\documentclass[manuscript]{aastex}
\usepackage{graphicx}

\begin{document}

\title{On the Rotation of the Magnetic Field Across the Heliopause}

\author{M. Opher\altaffilmark{1}}
\affil{Astronomy Department, Boston University, 725 Commonwealth
  Avenue, Boston, MA}
\email{mopher@bu.edu}

\author{J. F. Drake\altaffilmark{2}}
\affil{Department of Physics and the Institute for Physical Science and Technology, University of Maryland, College Park, MD} 
 
\begin{abstract} 
There was an expectation that the magnetic field direction would rotate dramatically across the heliopause (HP). This rotation was used as one of the criteria to determine if Voyager 1 (V1) had crossed the HP. Recently the Voyager team concluded that V1 crossed into interstellar space last year (Gurnett et al. 2013). The question is then why there was no significant rotation in the direction of the magnetic field across the HP (Burlaga et al. 2013). Here we present simulations that reveal that strong rotations in the direction of the magnetic field at the HP at the location of V1 (and Voyager 2) are not expected. The solar magnetic field strongly affects the drapping of the interstellar magnetic field ($B_{ISM}$) around the HP. $B_{ISM}$ twists as it approaches the HP and acquires a strong T component (east-west). The strong increase in the T component occurs where the interstellar flow stagnates in front of the HP. At this same location the N component $B_N$ is significantly reduced. Above and below the neighboring interstellar magnetic field lines also twist into the T direction. This behavior occurs for a wide range of orientations of $B_{ISM}$. Thus, the $B_{N}$ component at the V1 location and the associated angle $\delta=asin⁡(B_{N}/B)$ are small (around $10-20^{\circ}$), as seen in the observations (Burlaga et al. 2013). Only after some significant distance outside the HP, is the direction of the interstellar field distinguishably different from that of the Parker spiral. 
\end{abstract}


\section{Introduction}

Voyager 1 (V1) is at $125~AU$ from Earth travelling towards the nose of the heliosphere in the northern hemisphere. Voyager 2 (V2) is trailling behind at $102~AU$ travelling in the southern hemisphere.

Within the heliosheath (HS) the solar magnetic field  $B_{SW}$ is the Parker spiral with a dominant east-west orientation. The interstellar magnetic field ($B_{ISM}$) is expected to have a component in the north-south direction to account for the asymmetries in the heliosphere (Opher et al. 2006; 2007; 2009; Izmodenov et al. 2009; Pogorelov et al. 2009). Consequently, there is an expectation that the magnetic field direction will rotate dramatically across the heliopause (HP), which is the boundary that separates the plasma domain of the sun from that of the interstellar medium. This rotation was  used as one of the criteria to determine if V1 has already crossed the HP. Based on this fact, the absence of a significant rotation in the direction of the magnetic field at the times of dropouts of energetic particles produced within the heliosphere were interpreted as indicating that V1 was still in the heliosheath (Burlaga et al. 2013; Krimigis et al. 2013; Stone et al. 2013). However, simulations suggested a more complex HP with magnetic islands that would produce dropouts in the intensity of HS particles with essentially no local magnetic field rotation (Swisdak et al. 2013). In such a picture V1 might have already  crossed the HP just before the dropouts. Recently the Voyager team indeed concluded that V1 was in interstellar space based on the elevated plasma density inferred from plasma wave measurements (Gurnett et al. 2013). The crossing was conjectured to have happened around the time of the dropouts in August of 2012. On the other hand, others have suggestioned that V1 remains in the HS (Fisk \& Gloeckler 2013; McComas \& Schwadron 2012). In any case, if V1 is in interstellar space the important question is why V1 has not revealed a significant rotation in the direction of the magnetic field outside the HP (Burlaga et al. 2013). 

In this letter we present magnetohydrodynamic simulations of the global heliosphere that reveal that strong rotations in the direction of the magnetic field at the HP at the location of V1 (and V2) are not expected.  Only after some significant distance outside the HP, is the direction of the field distinguishably different from a Parker spiral. This results implies that the magnetic field orientation cannot be used as a marker for the crossing of the HP for the Voyager spacecrafts.

In the next section we describe the twist of the interstellar magnetic field as it approaches the HP and then make some concluding remarks.

\section{Twist of the Interstellar Magnetic Field}

To study the twist of the interstellar magnetic field we use our 3D MHD model (Opher et al. 2009), which is based on a multi-fluid description that includes adaptive mesh refinement as well as the magnetic field of the sun and the interstellar magnetic field ($B_{ISM}$). It possesses 5-fluids, 1 ionized and 4 neutral H fluids. The multi-fluid approach for the neutrals (Alexashov \& Izmodenov 2005, Zank 1999) captures the main features of the kinetic model (Izmodenov 2009). Atoms of interstellar origin represent population 4. Population 1 appears in the region behind the bow shock (or slow shock, depending on the intensity of $B_{ISM}$ (Zieger et al. 2013). Populations 3 and 2 appear in the supersonic solar wind and in the compressed region behind the TS, respectively. All four populations are described by separate systems of the Euler equations with corresponding source terms describing neutral-ion charge exchange.

The inner boundary of our domain is a sphere at $30~A$U and the outer boundary is at $x = \pm 1000~AU$, $y = \pm 1000~AU$, $z = \pm 1000~AU$. Parameters of the solar wind at the inner boundary were chosen to match the values obtained by (Izmodenov 2009) at $30~AU$: $V_{SW} = 417~km/s$, $n_{SW} = 8.74 \times 10^{-3}~ cm^{-3}$, $T_{SW} = 1.087\times 10^{5}~K$ and the Parker spiral magnetic field $B_{SW} = 7.17\times 10^{-3}~nT$ at the equator. In our simulation, we assume that the magnetic axis is aligned with the solar rotation axis. The solar wind flow at the inner boundary is assumed to be spherically symmetric. For the interstellar plasma we assume: $v_{ISM} = 26.4 km/s$, $n_{ISM} = 0.06~ cm^{-3}$, $T_{ISM} = 6519~ K$. The number density of H atoms in the interstellar medium is $n_{H} = 0.18 cm^{-3}$, the velocity and temperature are the same as for the interstellar plasma. The coordinate system is such that Z-axis is parallel to the solar rotation axis, X-axis is $5^{\circ}$ above the direction of interstellar flow with Y completing the right-handed coordinate system (a schematic figure can be found in Alouani et al. (2011)). The grid domain has 14 million cells ranging from scales of $0.24~AU$ at the inner boundary and $1.0~AU$ (for cases (a)-(b) in Figure 1) and $2.0~ AU$ (for case (c) in Figure 1) at the HP. 

The strength of the $B_{ISM}$ in the model is $4.4\mu G$. The orientation of $B_{ISM}$ continues to be debated in the literature. The orientation of $B_{ISM}$ is defined by two angles, $\alpha_{BV}$ and  $\beta_{BV}$. ($\alpha_{BV}$ is the angle between the interstellar magnetic field and flow velocity of the interstellar wind and $\beta_{BV}$ is the angle between the $B_{ISM}-v_{ISM}$ plane and the solar heliographic equator).  In studies like (Opher et al. 2009; Izmodenov et al. 2009) small values of $\alpha_{BV} \approx 10^{\circ}- 20^{\circ}$ were required to account for the heliospheric asymmetries, such as the different crossing distances of the termination shock by V1 and 2 (Stone et al. 2008). Others (McComas et al. 2009; Heerikhuisen \& Pogorelov 2011; Chalov et al. 2010) have used the observed shape and location of the IBEX ribbon to constrain the magnitude and orientation of $B_{ISM}$. However, such constraints are sensitive to the specific model of the IBEX ribbon, which continues to be uncertain. Because of the uncertainties associated with the modeling of the IBEX ribbon, we take a strong $B_{ISM}$ ($4.4 \mu G$) and an orientation that accounts for the heliospheric asymmetries (Opher et al. 2009). In any case, as we show that the twist of the interstellar magnetic field just outside of the HP is insensitive to its original orientation. 

As illustrated in Figure 1 there is a pile-up of the tangential component of the interstellar magnetic field, $B_{T}$, outside of the HP that is independent of the original orientation of $B_{ISM}$. (Here we use the R-T-N coordinate system as the local cartesian system centered in the Sun. R is radially outward from the Sun, T is in the plane of the solar equator and positive in the direction of solar rotation, and N completes a right-handed system). The pile-up of $B_{T}$ holds true as well for cases, where we included a latitudinal varying solar wind (Figure 5a) or a dipole magnetic configuration for the solar wind. 

\begin{figure}[htbp]
\centering
\includegraphics[width=0.3\textwidth]{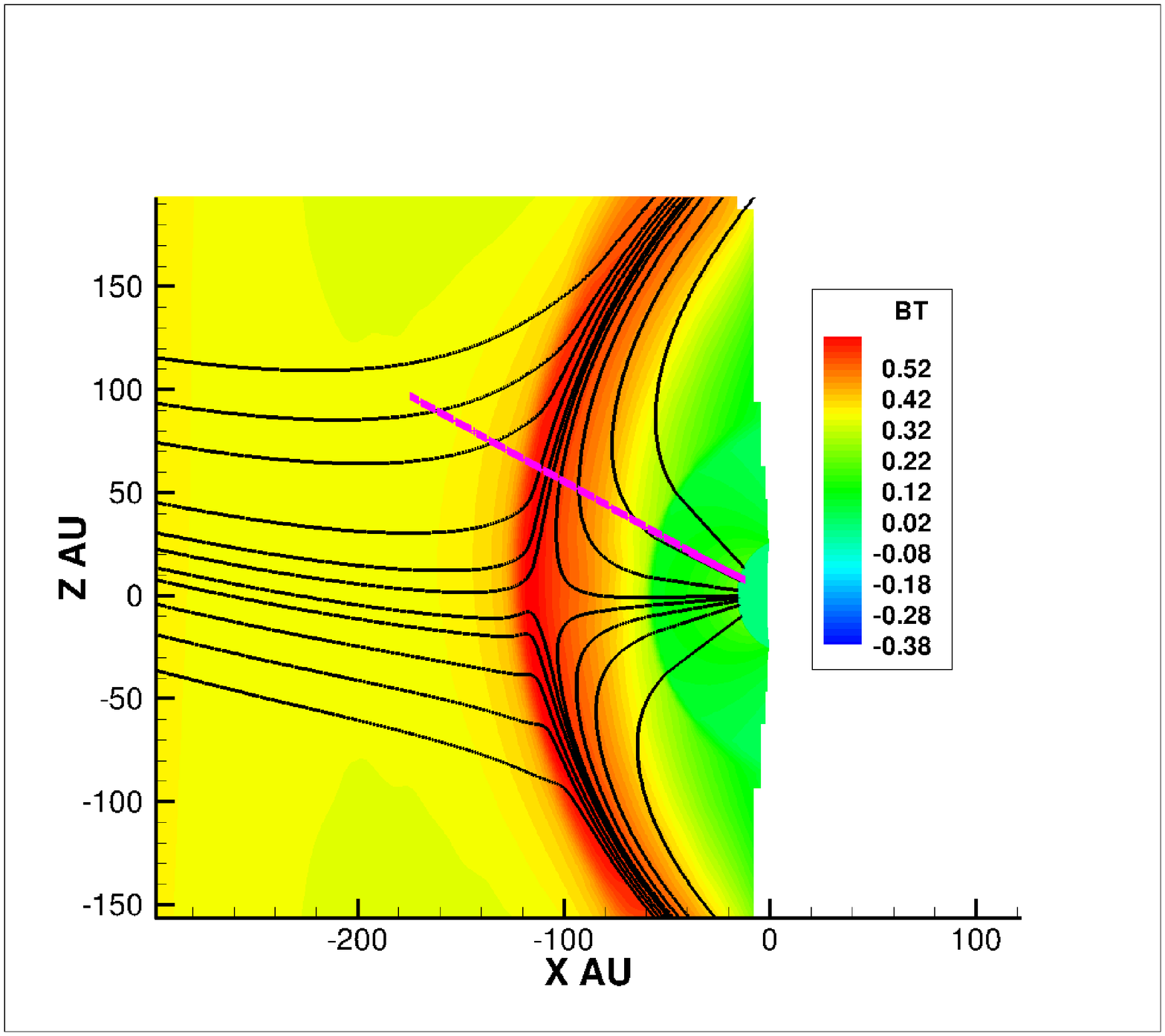} 
 \includegraphics[width=0.3\textwidth]{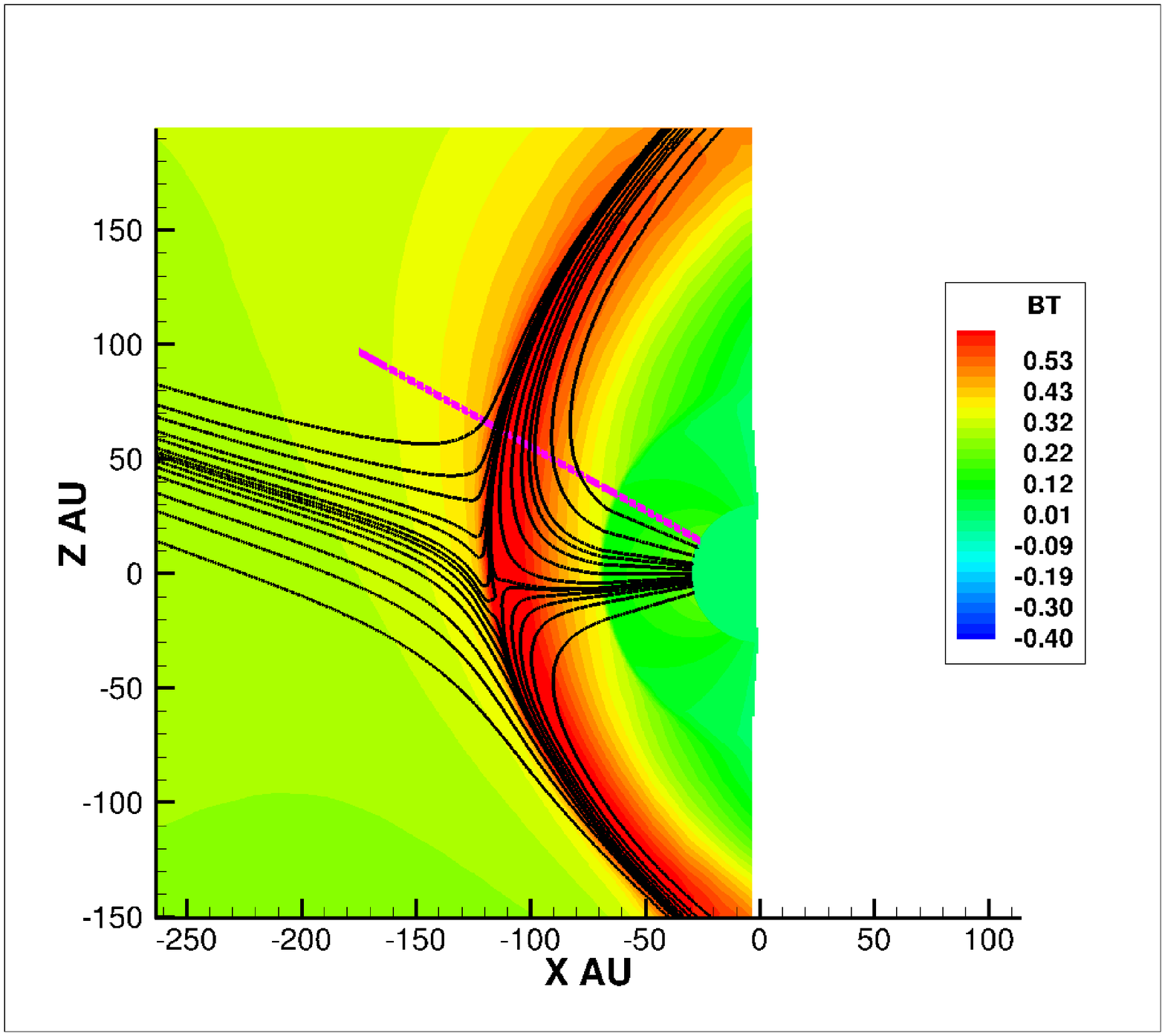}  
 \includegraphics[width=0.3\textwidth]{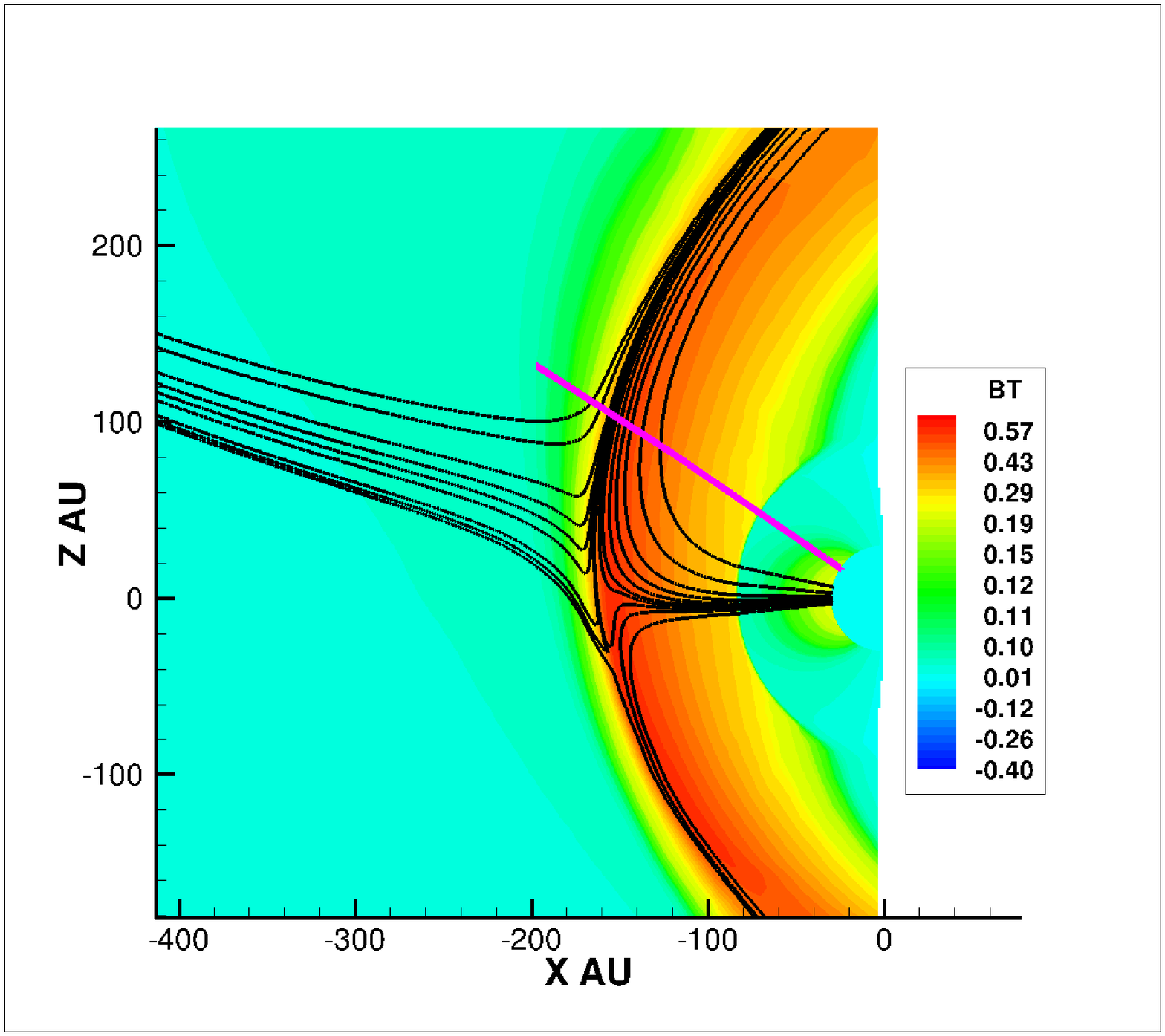}          
\caption{Pile up of the tangential component outside of the HP. The tangential component BT, of $B_{ISM}$ in the Voyager 1-x plane for different orientations of $B_{ISM}$: (a) $\beta_{BV}=51.5^{\circ}$; $\alpha_{BV}=15.9^{\circ}$ (b) $\beta_{BV}=57.3^{\circ}$; $\alpha_{BV}=40.7^{\circ}$, and (c) $\beta_{BV}=76.9^{\circ}$; $\alpha_{BV}=15.1^{\circ}$. All these cases assume a monopole magnetic field and a uniform speed for the solar wind. The monopole field was chosen to reduce the artificial numerical magnetic reconnection that takes place at the HP compared with a dipolar solar field model. The black lines are flow streamlines. The HP can be identified as the location where the flow streamlines from the ISM encounter those of the solar wind. The magenta line marks the Voyager 1 trajectory. $\alpha_{BV}$ is the angle between the pristine $B_{ISM}$ and the interstellar velocity $v_{ISM}$ of the interstellar wind. $\beta_{BV}$ is the angle between the $B_{ISM}-v_{ISM}$ plane and the solar heliographic equator. The R-T-N coordinate system is the local cartesian system centered in the Sun. R is radially outward from the Sun, T is in the plane of the solar equator and is positive in the direction of solar rotation, and N completes a right-handed system.}
\label{figure1}
\end{figure}

The angle $\delta=asin⁡(B_{N}/B)$ is a measure of how much the magnetic field orientation deviates from a Parker spiral. Inside the heliosphere when the solar magnetic field is mainly in the T direction $\delta \approx 0^{\circ}$ (although it can acquire a finite but small value close to the HP). One can see that for different orientations of $B_{ISM}$, there is a layer just outside the HP where the angle $\delta$ remains small (around $14^{\circ}-20^{\circ}$) as measured by V1 (Burlaga et al. 2013) (Figure 2). At Voyager 2 (V2) the angle $\delta$ is also small just ahead of the HP even for a large value of $\alpha_{BV}$ (panel (b)).

\begin{figure}[htbp]
\centering
\includegraphics[width=0.3\textwidth]{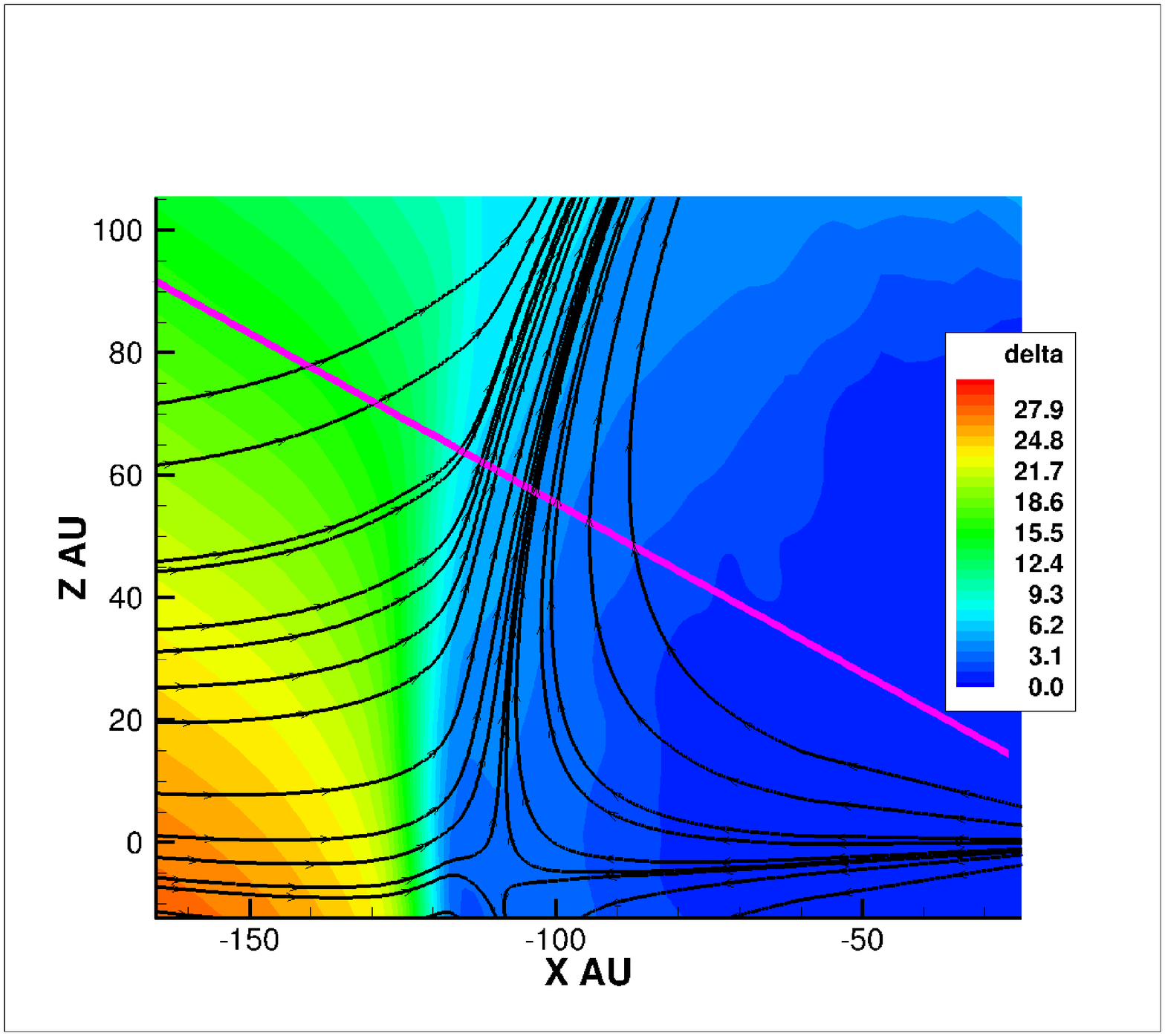}          
\includegraphics[width=0.3\textwidth]{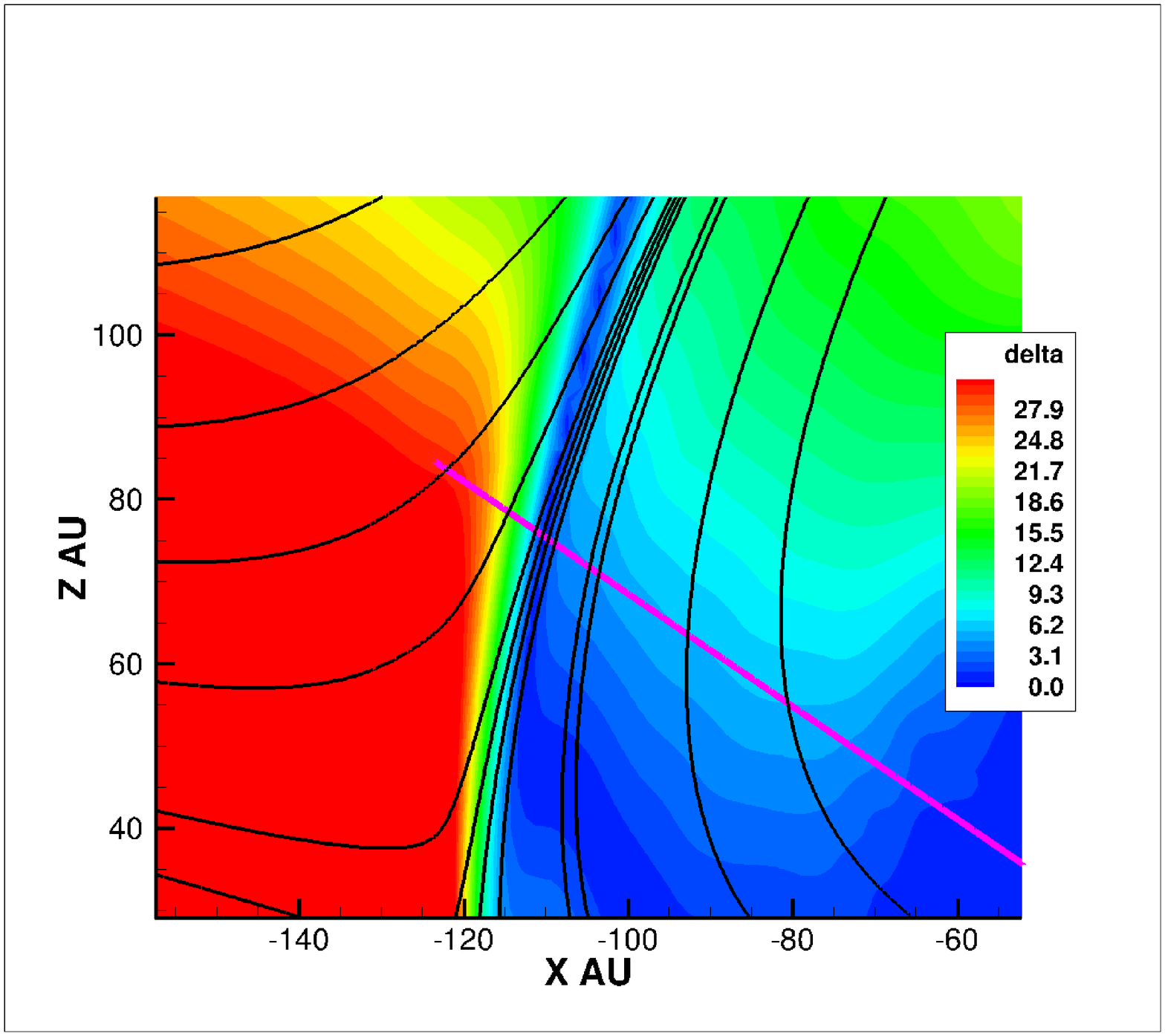}     
\includegraphics[width=0.3\textwidth]{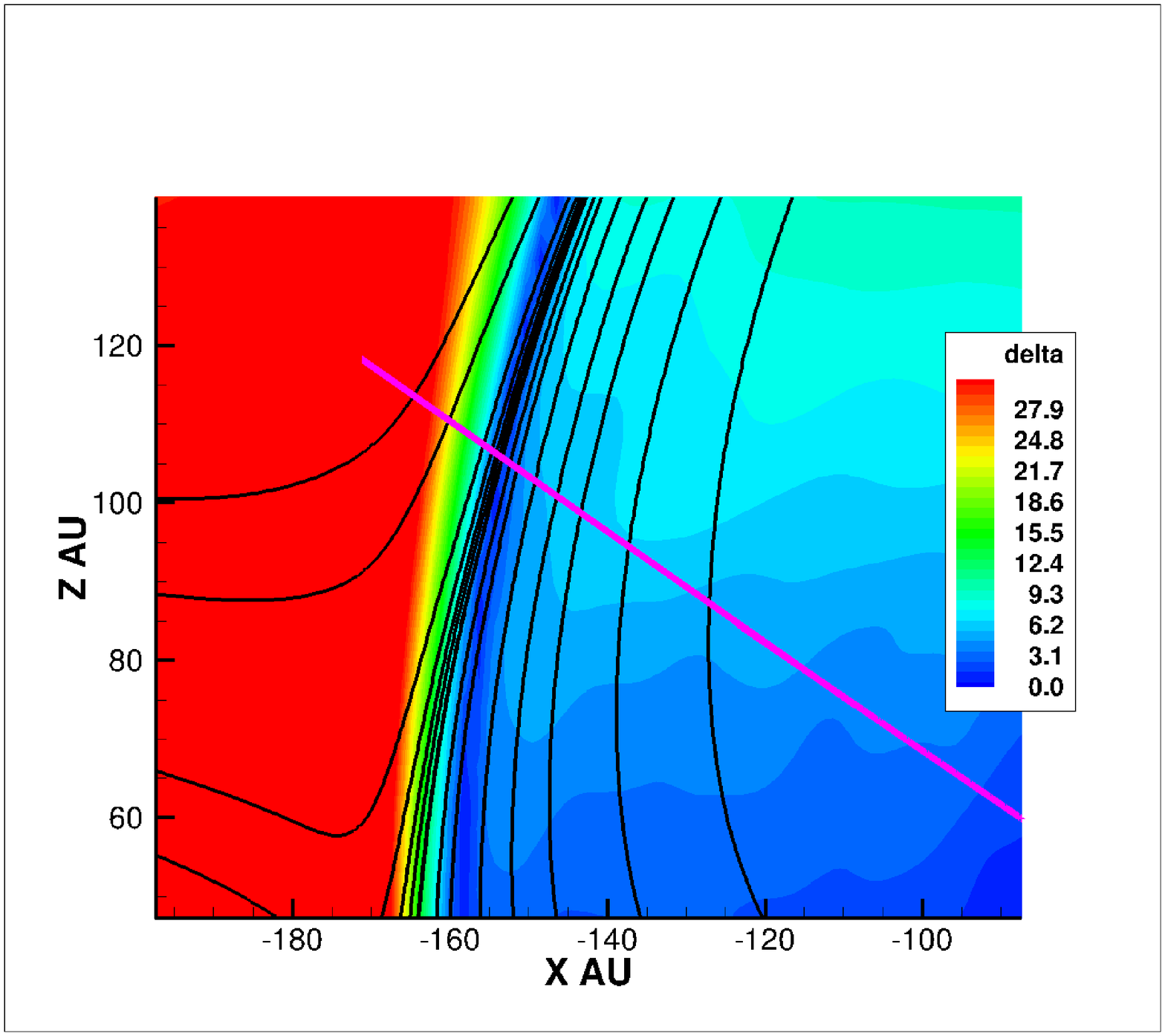}          
\includegraphics[width=0.3\textwidth]{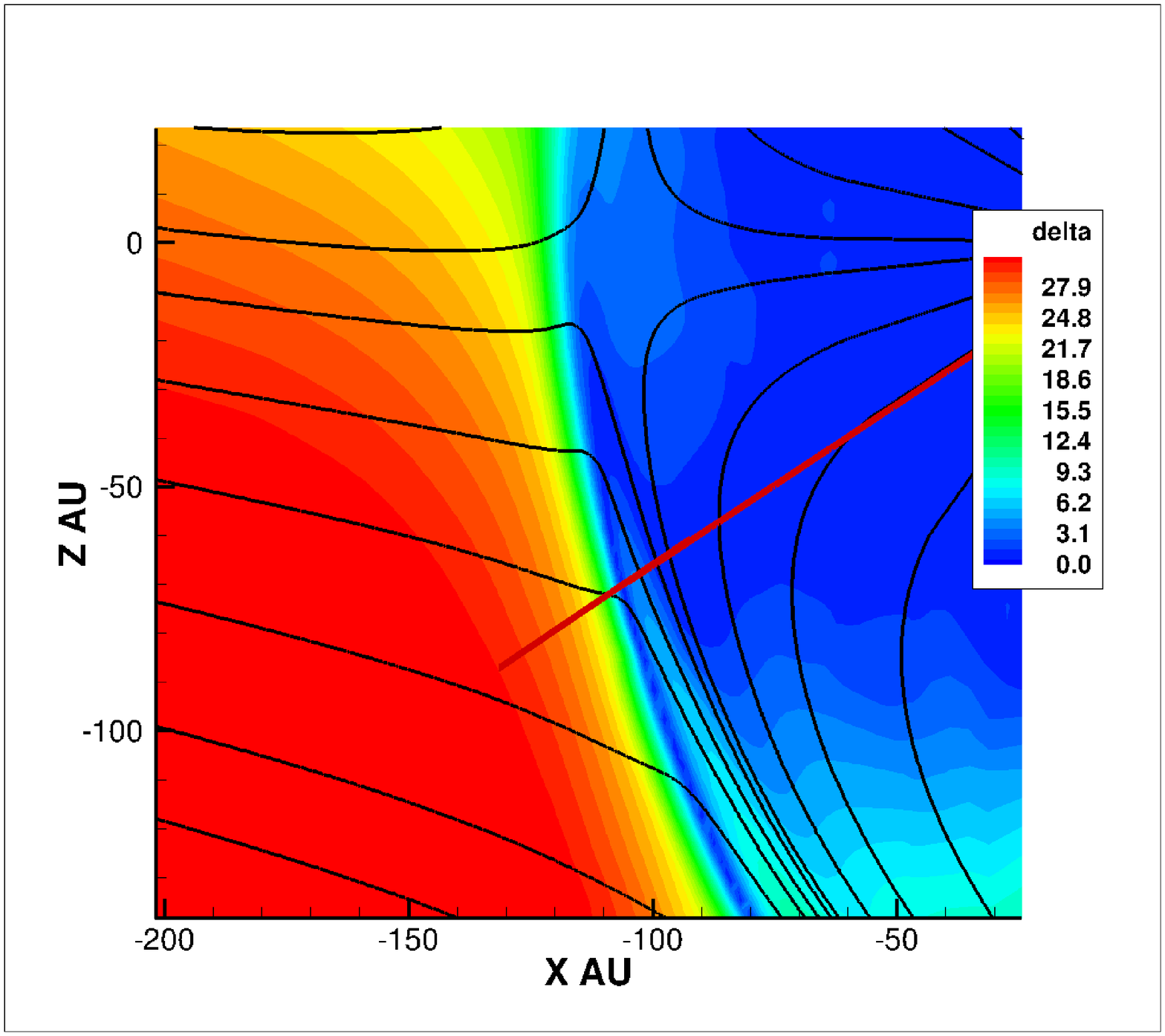}     
\includegraphics[width=0.3\textwidth]{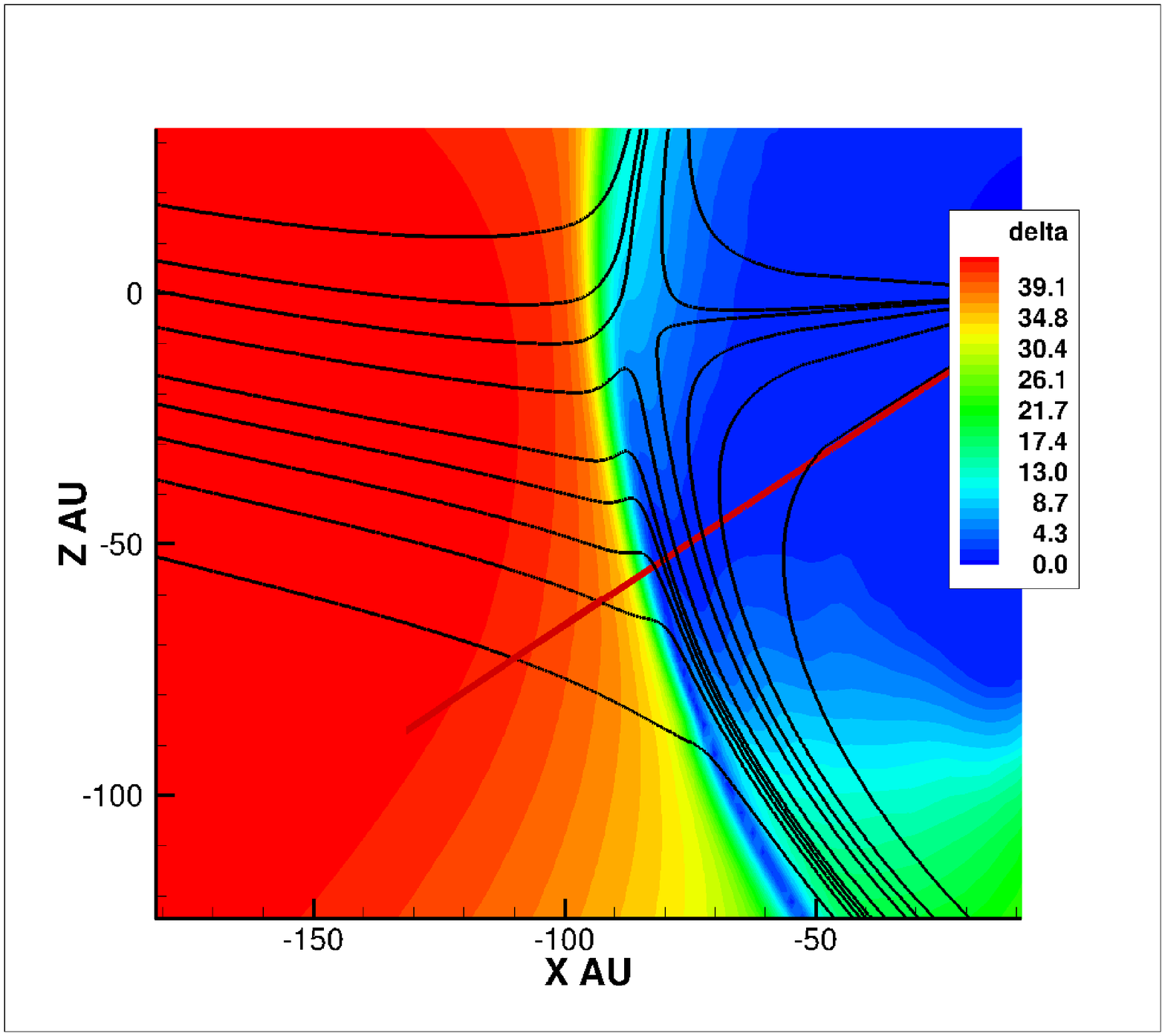}     
\includegraphics[width=0.3\textwidth]{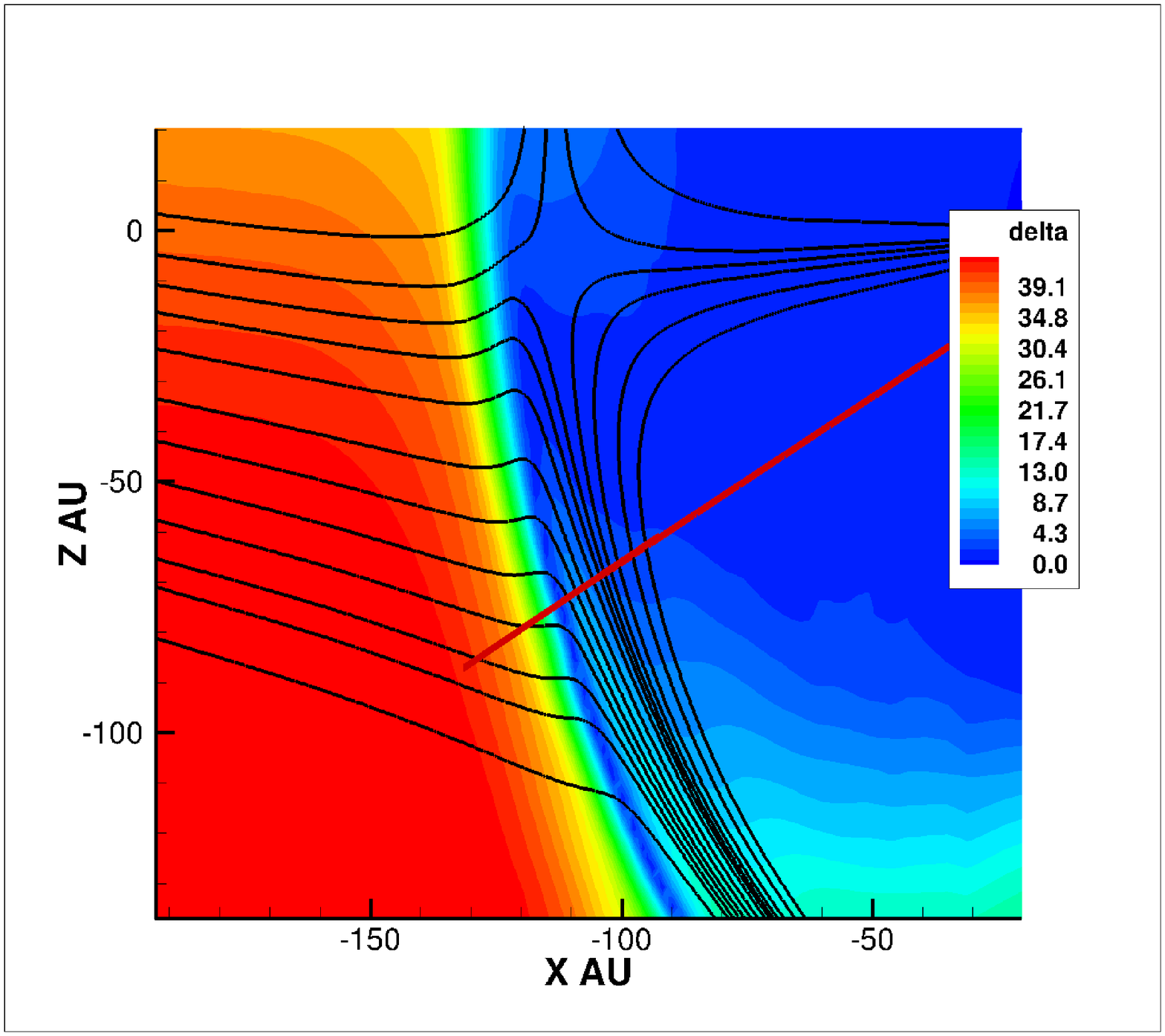}     
\caption{The angle $\delta$ outside of the HP. The angle $\delta=asin⁡(B_{N}/B)$ (in $^{\circ}$) at the V1-x plane for the same  configurations of $B_{ISM}$ as in Figure 1. The magenta line indicates the Voyager 1 trajectory. Panels (d-g) show the same as panels (a-c) but for V2-x plane. The red line indicates the Voyager 2 trajectory.}
\label{figure2}
\end{figure}

The $B_{ISM}$ therefore undergoes a strong twist just before reaching the HP and aligns itself mainly in the T direction. This can be seen in Figure 3 (a-b), where at large distances outside of the HP the interstellar field lines are inclined to the T direction (east-west direction) and then twist dramatically in the T direction as they approach the HP. This twist does not take place when the solar magnetic field $B_{SW}$ is absent (Figures 3c-d). In this case the interstellar magnetic field lines remain largely in their original orientation with respect to the BV plane (the plane formed by the $B_{ISM}$ and $v_{ISM}$ vectors). As the field lines approach the HP they drape around the HP in a symmetric manner with respect to the BV plane rather than by twisting into the direction of the solar spiral magnetic field direction. As a consequence the HP becomes distinctly elongated in the direction of $B_{ISM}$  (Figure 3d). The shape of the HP in the case with $B_{SW}$  (Figures 3a-b) is in comparison much blunter and oriented towards the rotation axis of the Sun (N or z-axis). The solar magnetic field, therefore, plays a crucial role in controlling the shape and draping of the $B_{ISM}$ at the HP.

\begin{figure}[htbp]
\centering
\includegraphics[width=0.4\textwidth]{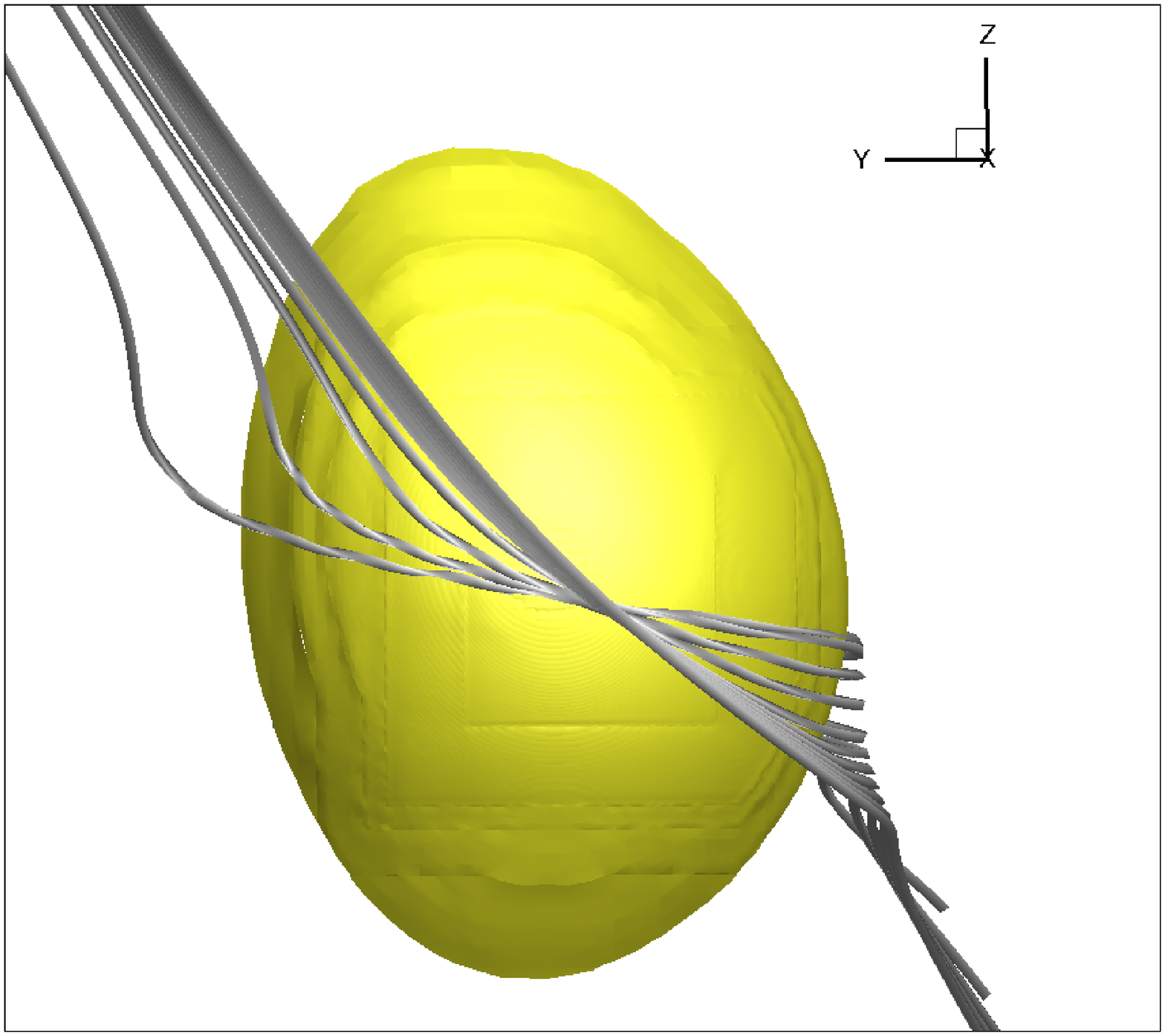}          
\includegraphics[width=0.4\textwidth]{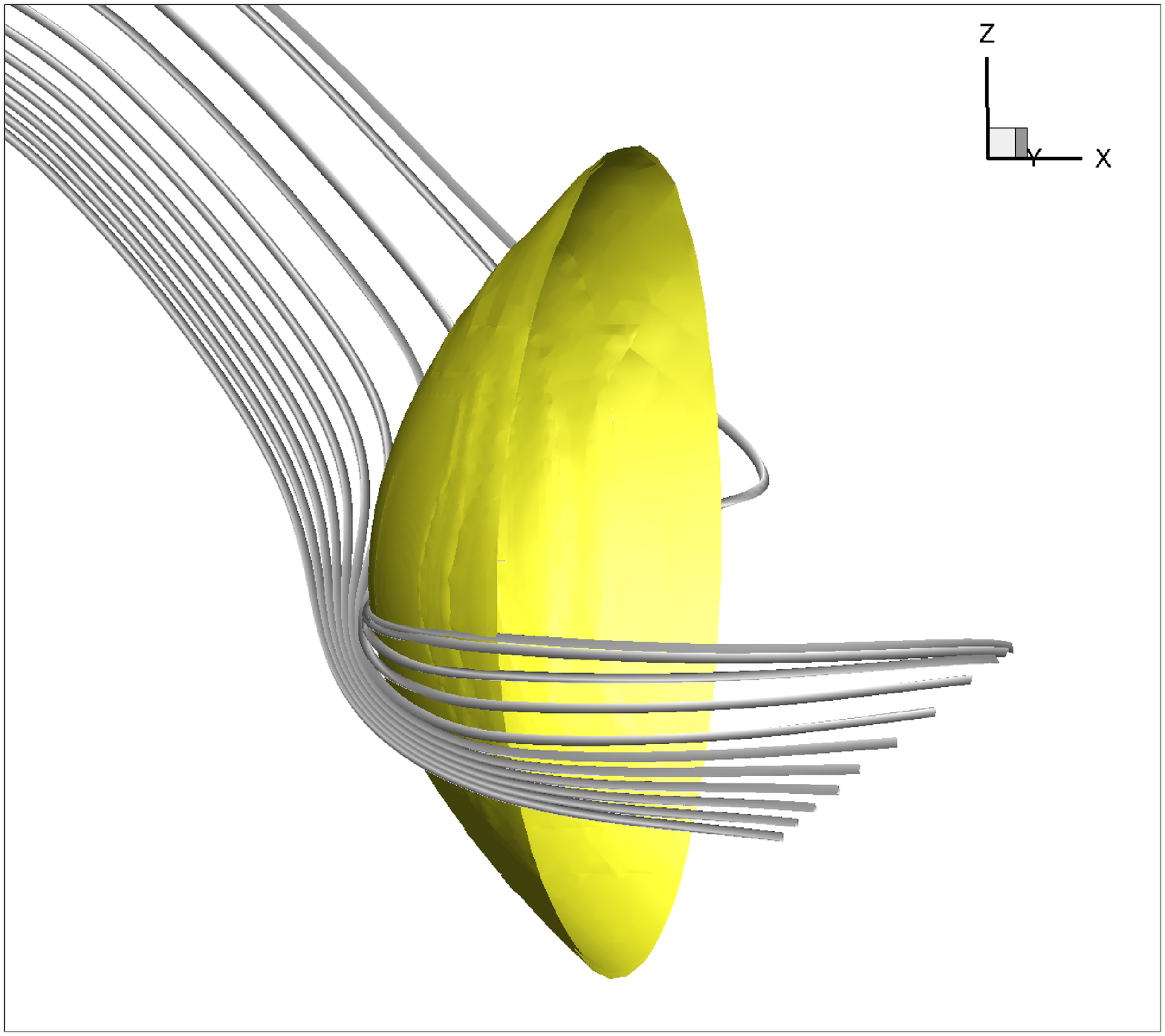}     
\includegraphics[width=0.4\textwidth]{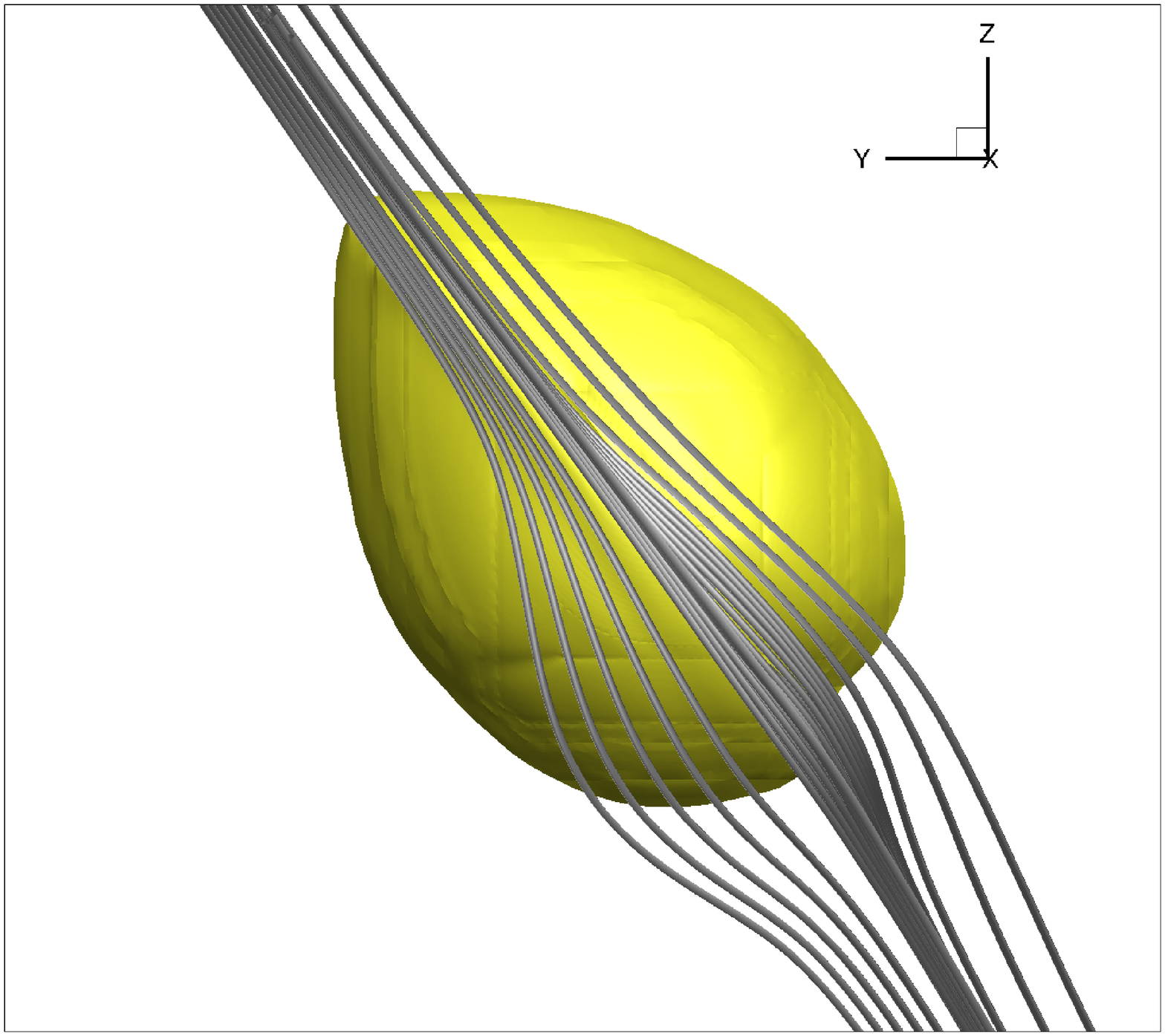}          
\includegraphics[width=0.4\textwidth]{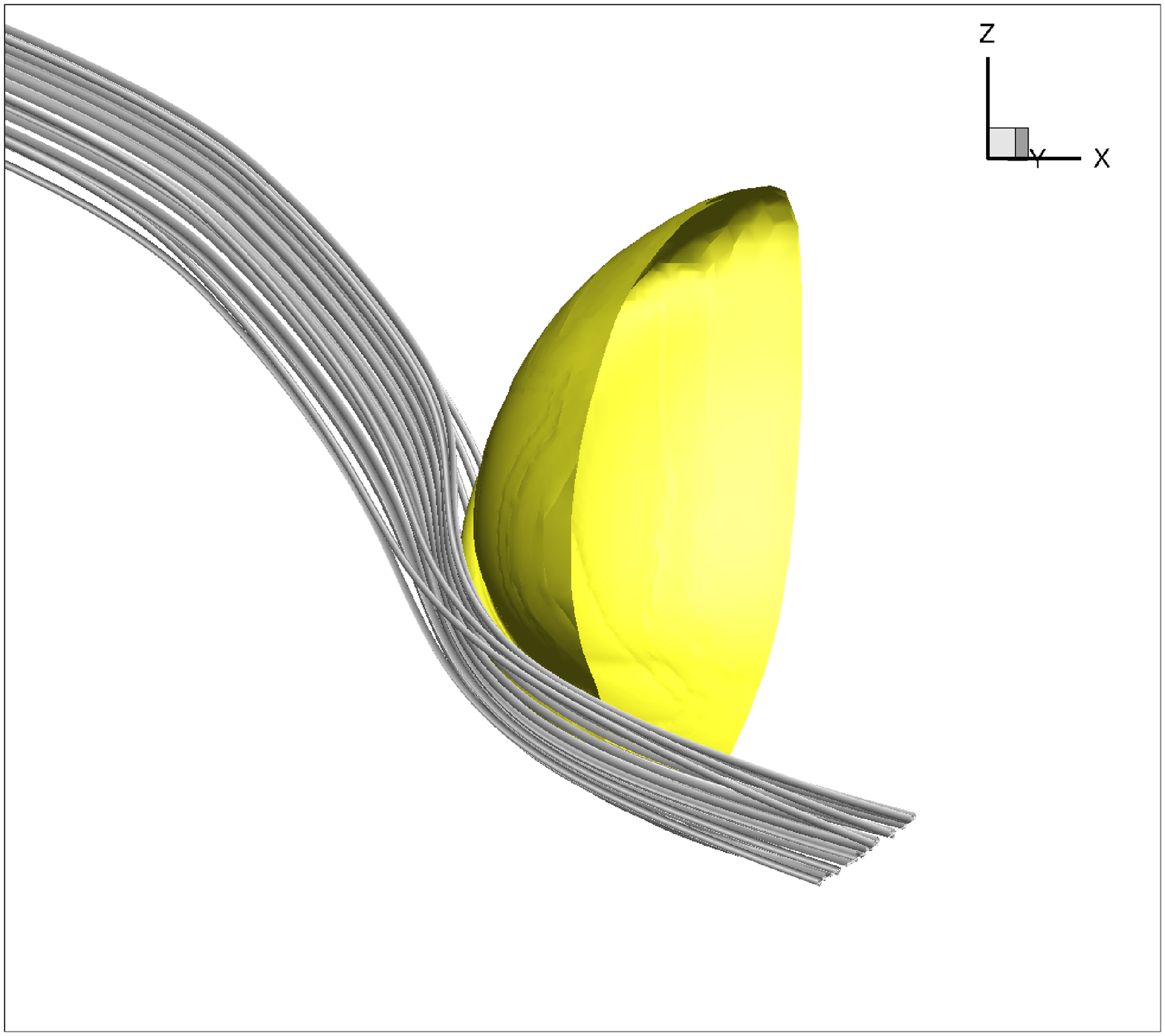}     
\caption{The twist of the interstellar magnetic field outside of the HP. View at the nose of the heliosphere from the interstellar medium towards the Sun for (a) $\beta_{BV}=51.5^{\circ}$; $\alpha_{BV}=15.9^{\circ}$; and in (b) a view from the side. The nose of the HP is shown in the yellow iso-surface (defined by $log T=11.9-12$). The gray field lines are the $B_{ISM}$ wrapping and twisting around the HP. Panels (c) and (d) are for same direction of $B_{ISM}$ as in panels (a) and (b) but for a simulation without a solar magnetic field.}
\label{figure3}
\end{figure}

Overall, the interstellar magnetic field slips around the heliosphere much more easily in the absence of $B_{SW}$ than in the case with $B_{SW}$. This is documented in Figure 4 where the magnetic fields and flows are compared with and without $B_{SW}$. First, the HP is much further from the sun without $B_{SW}$. Second, the interstellar magnetic field magnetic field piles up outside of the HP much more strongly with $B_{SW}$ than without $B_{SW}$. The peak values of BT are three times larger in Figure 4a compared with Figure 4d. Third, the flows $V_{N}$ (Figures 4c, 4f) are essentially discontinuous across the HP in the case without $B_{SW}$ while they are essentially continuous with $B_{SW}$. With $B_{SW}$ the $B_{ISM}$ therefore twists in the direction of $B_{SW}$ at the stagnation region. The field lines get “hung-up” in the stagnation region so that the magnetic field strength increases and exerts more pressure on the heliosphere than without $B_{SW}$, resulting in a smaller heliosphere. The normal flows $V_{N}$ just outside of the HP are also reduced as the magnetic field gets “hung-up” in the stagnation region (Figures 4c and f). The neighboring interstellar magnetic field lines above and below the stagnation region also twist in response to the magnetic field pile-up near the stagnation region. There is therefore a layer of strong $B_{T}$ and small $B_{N}$ outside of the HP with a finite latitudinal extent. In this region the angle $\delta$ is reduced (Figures 2 and 5c).

\begin{figure}[htbp]
\centering
\includegraphics[width=0.3\textwidth]{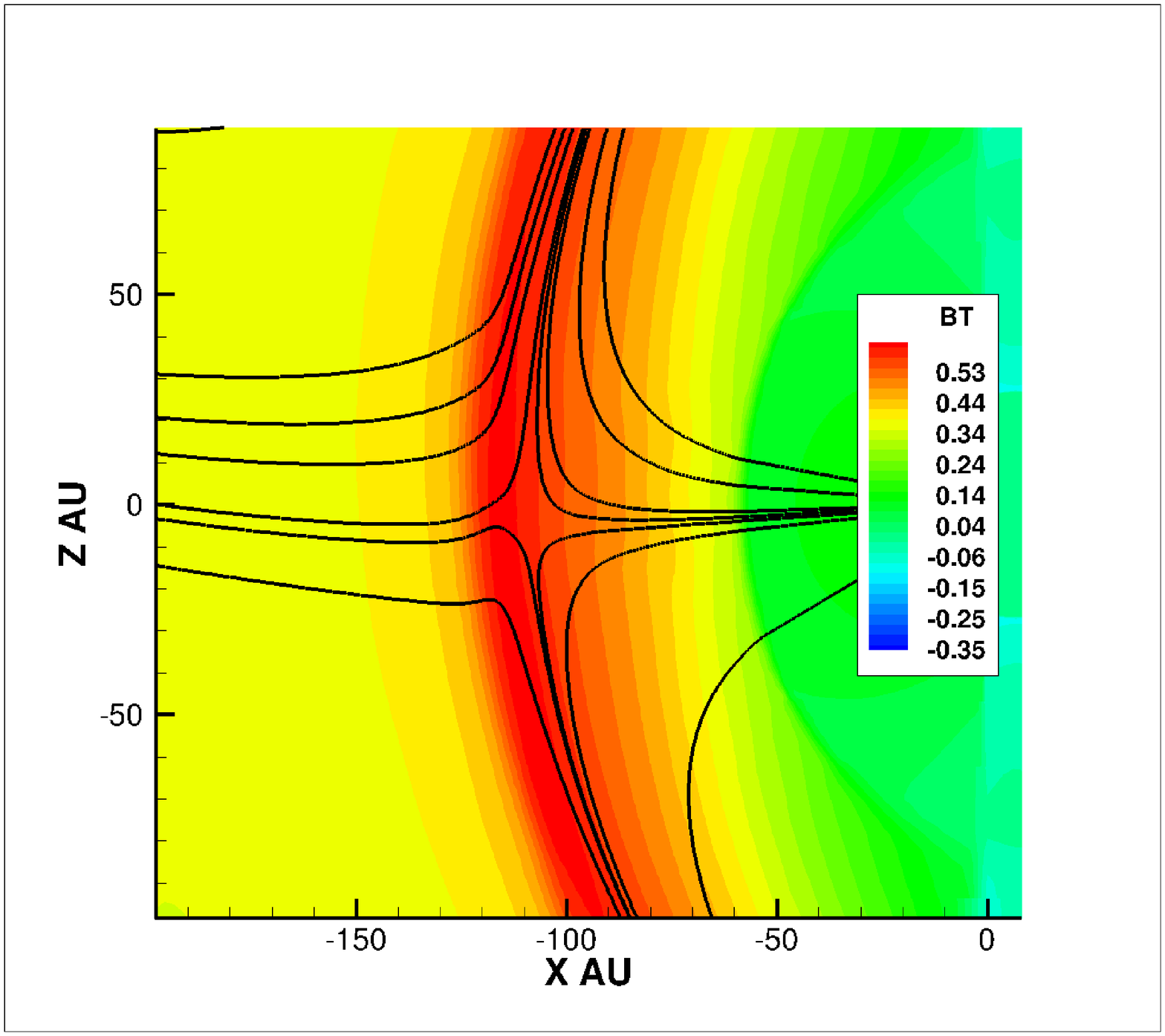}          
\includegraphics[width=0.3\textwidth]{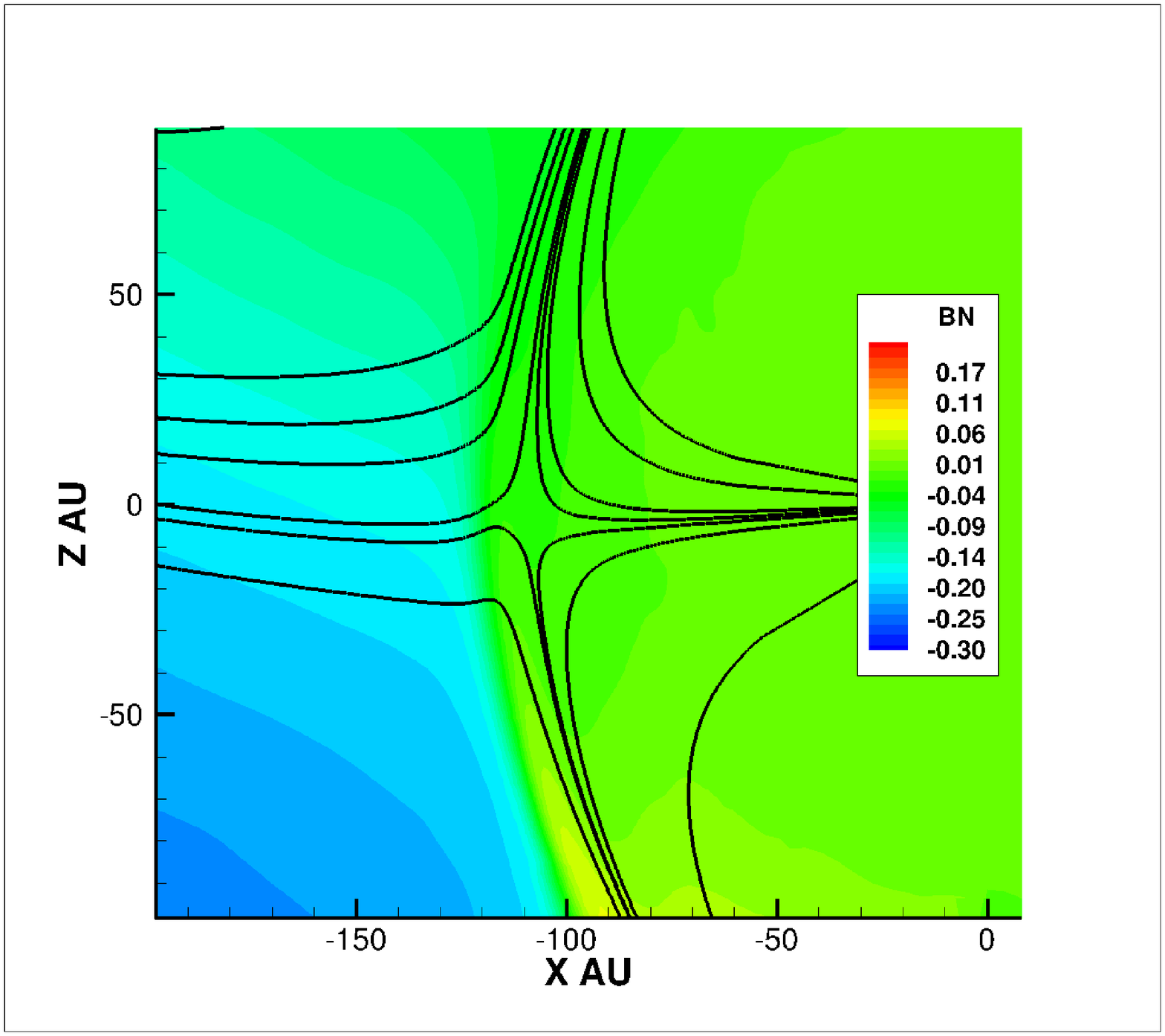}     
\includegraphics[width=0.3\textwidth]{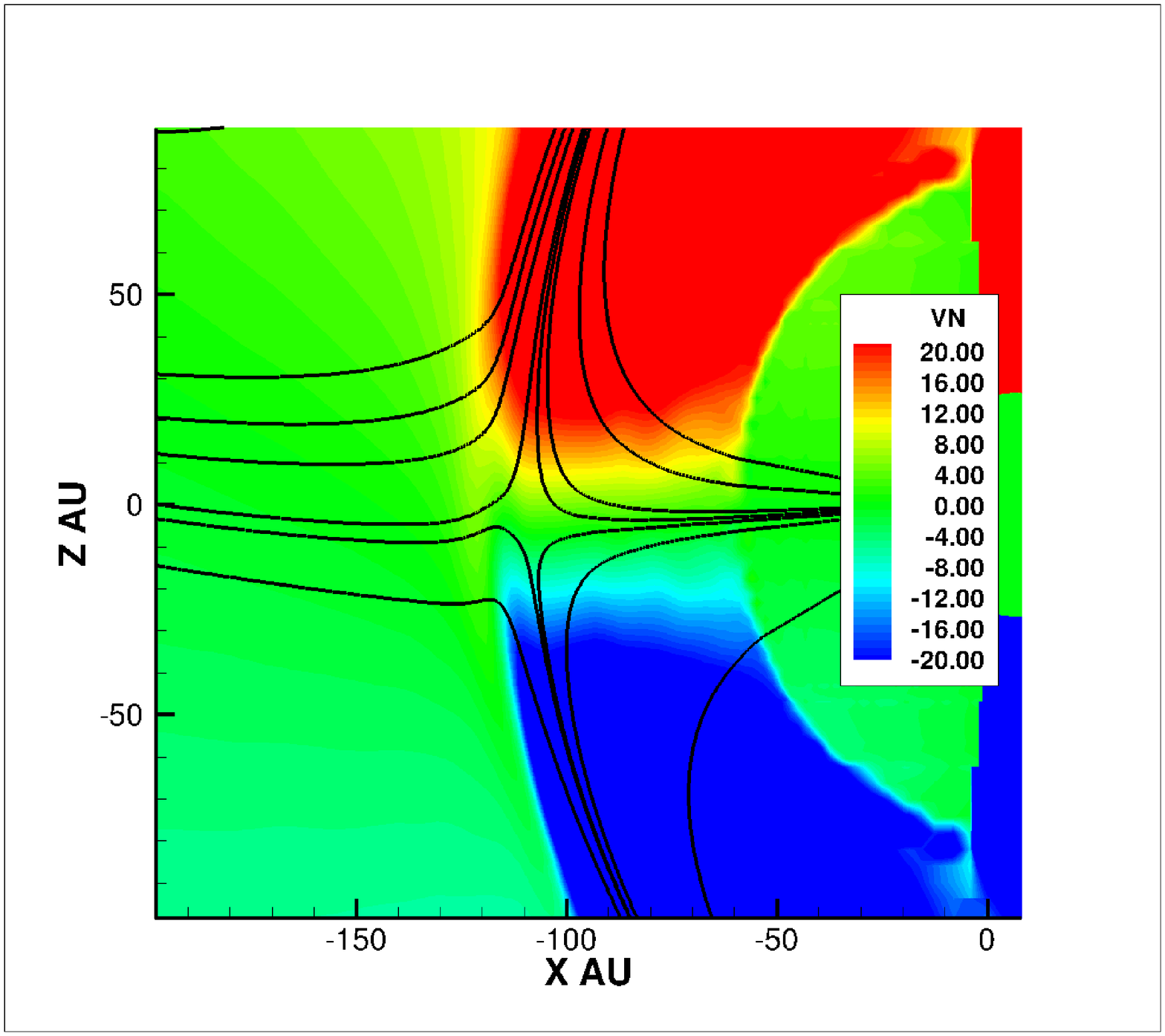}          
\includegraphics[width=0.3\textwidth]{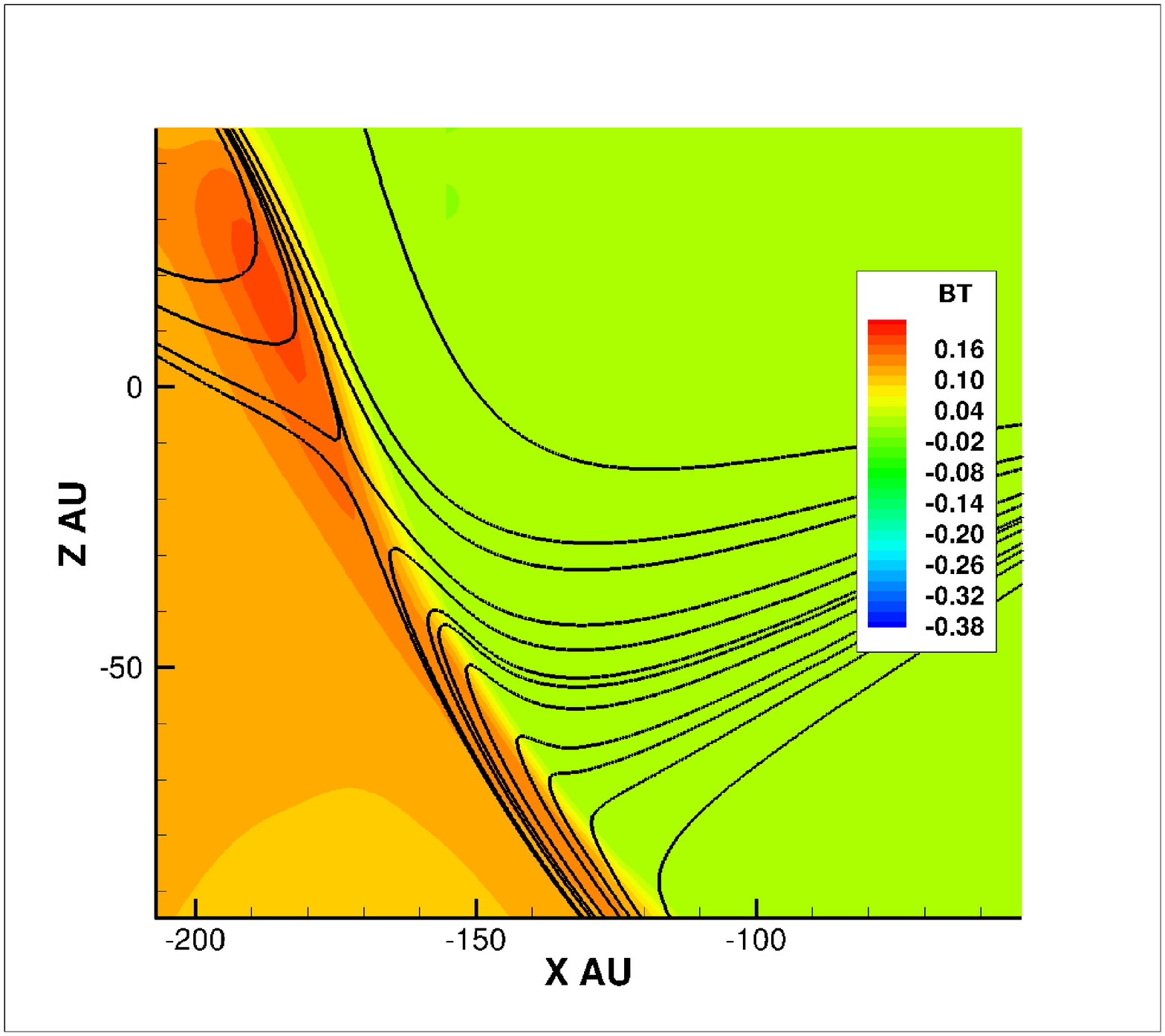}     
\includegraphics[width=0.3\textwidth]{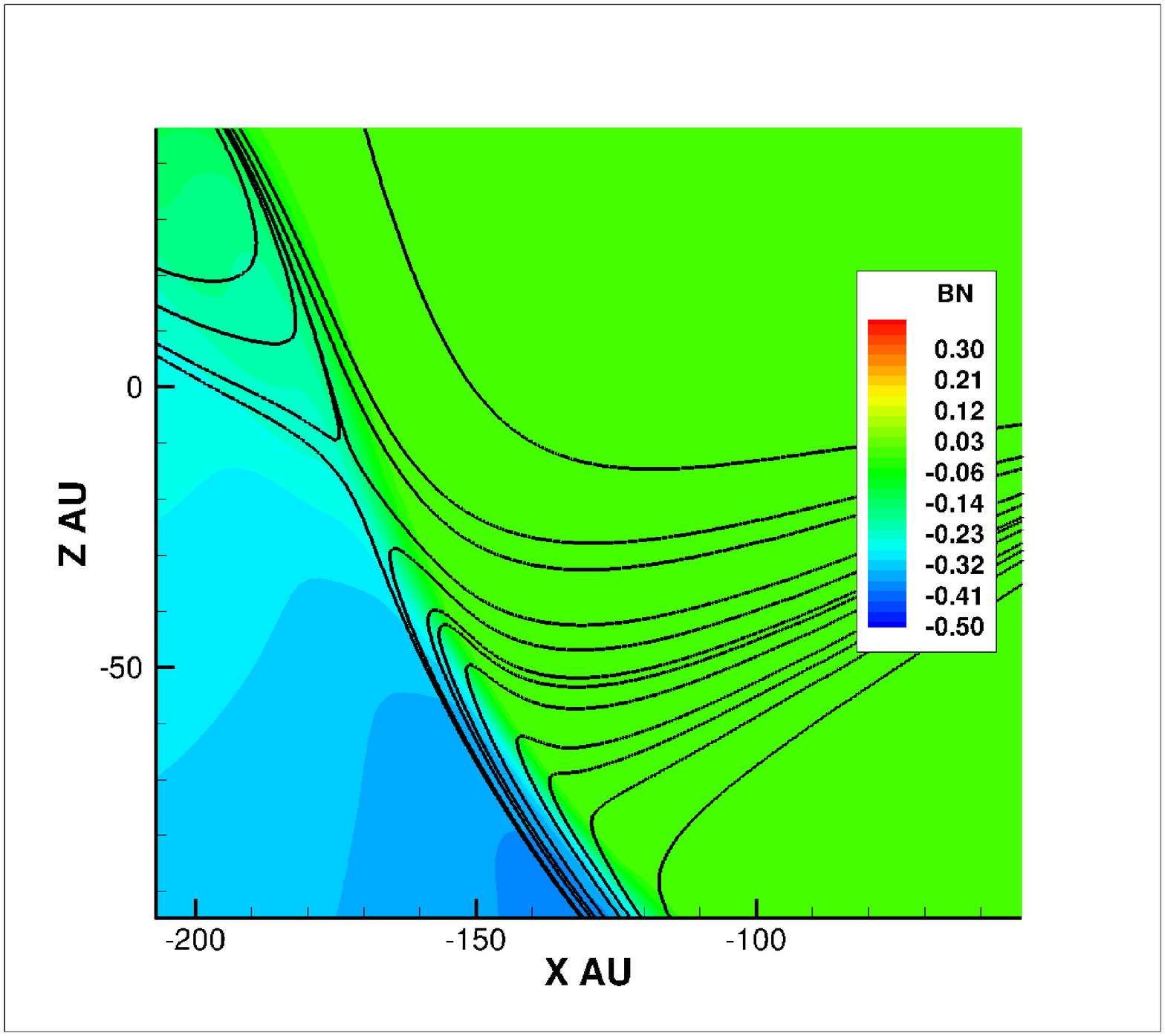}     
\includegraphics[width=0.3\textwidth]{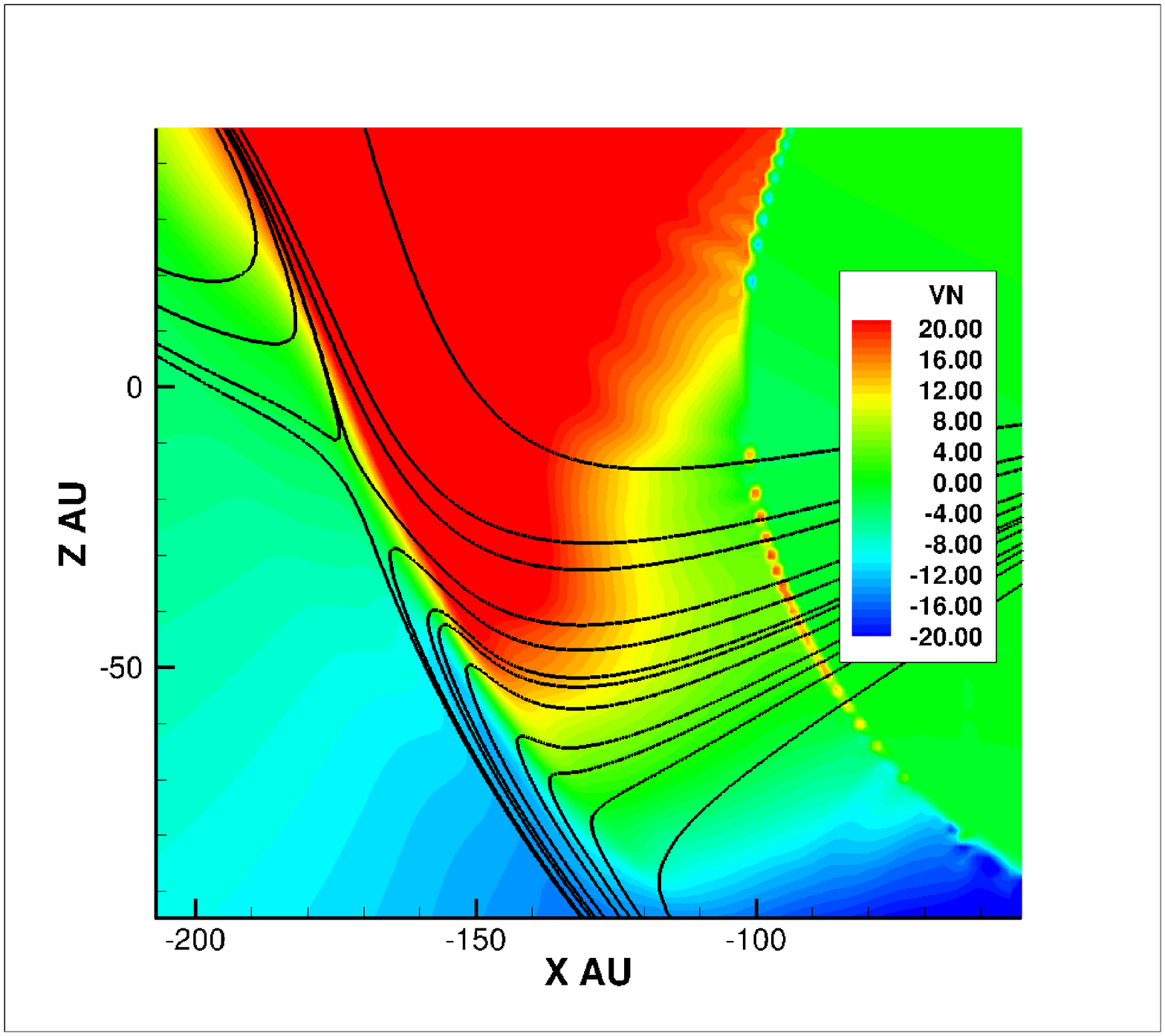}     
\caption{The behavior of $B_{ISM}$ and plasma flows near the Stagnation Point. $B_{T}$ and $B_{N}$ (nT) components in the V1-x plane (a-b) $\beta_{BV}=51.5^{\circ}$; $\alpha_{BV}=15.9^{\circ}$; with a monopole $B_{SW}$. Panels (d-e) are the same orientation of $B_{ISM}$ as in panels (a-b) but for a simulation with no $B_{SW}$. Panels (c) and (f) show $V_{N}(km/s)$ It can be seen that the normal flows VN outside in the interstellar medium, outside the HP are much reduced at the stagnation region for the case with $B_{SW}$.}
\label{figure4}
\end{figure}

\begin{figure}[htbp]
\centering       
\includegraphics[width=0.3\textwidth]{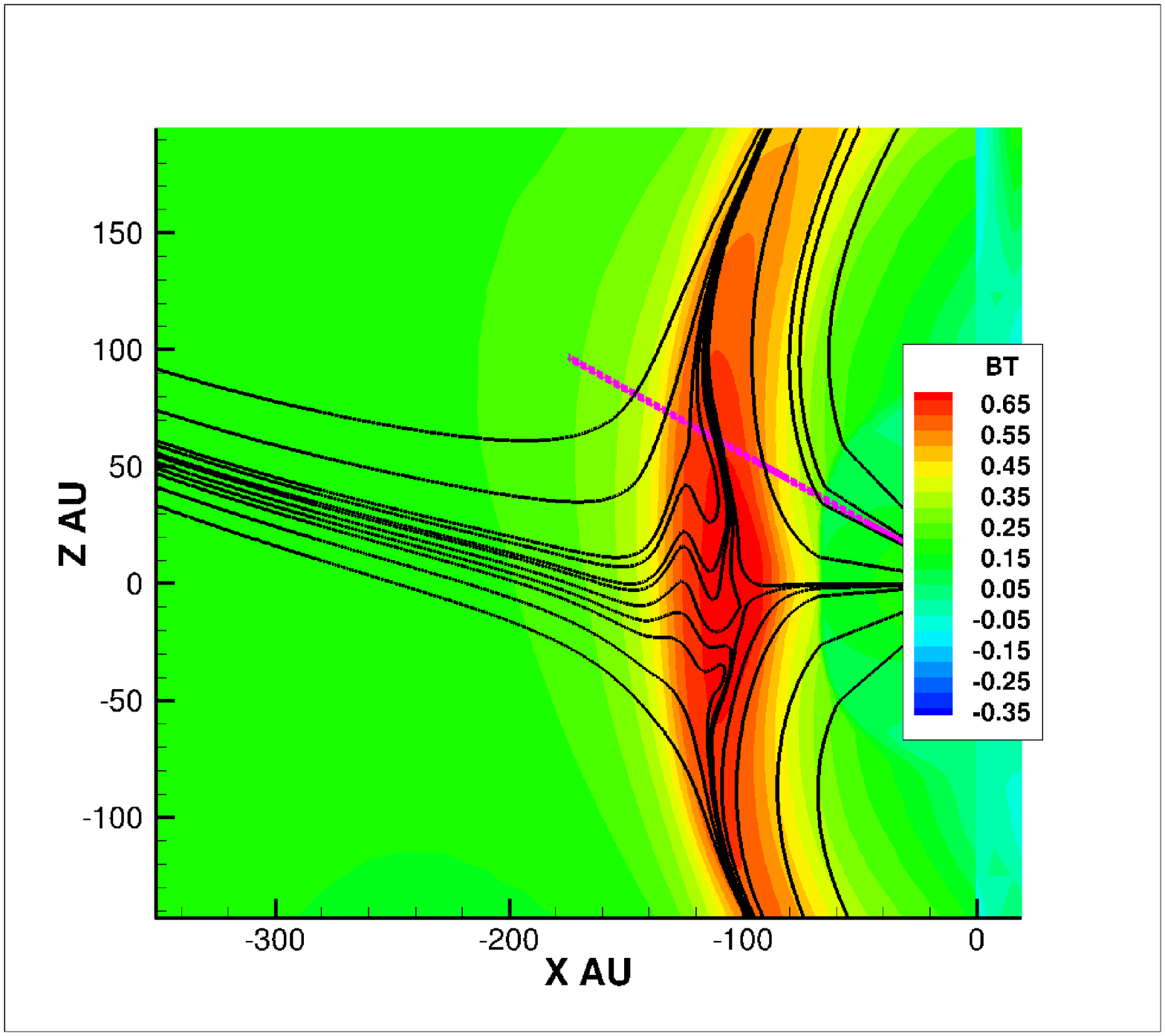}     
\includegraphics[width=0.3\textwidth]{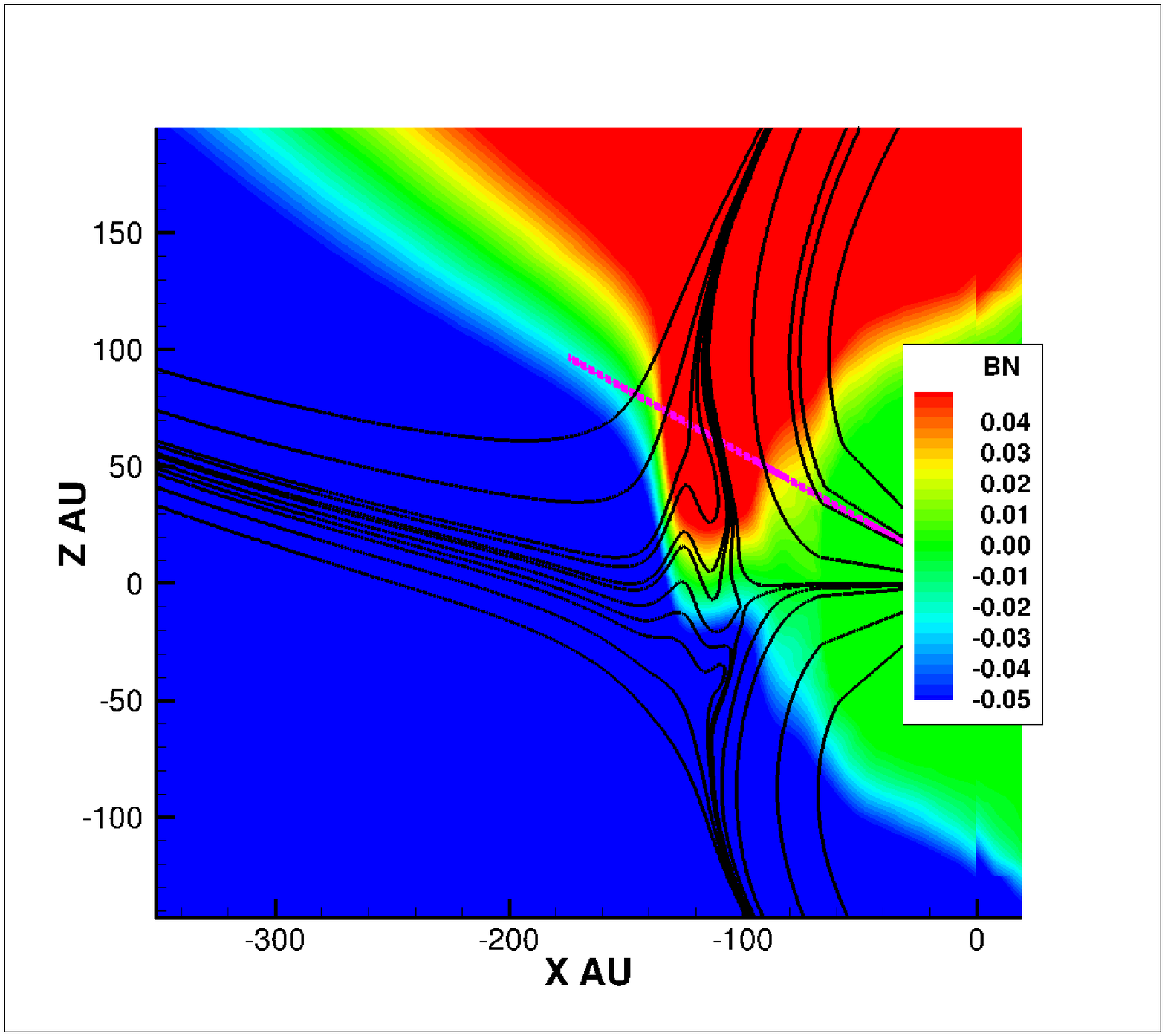}          
\includegraphics[width=0.3\textwidth]{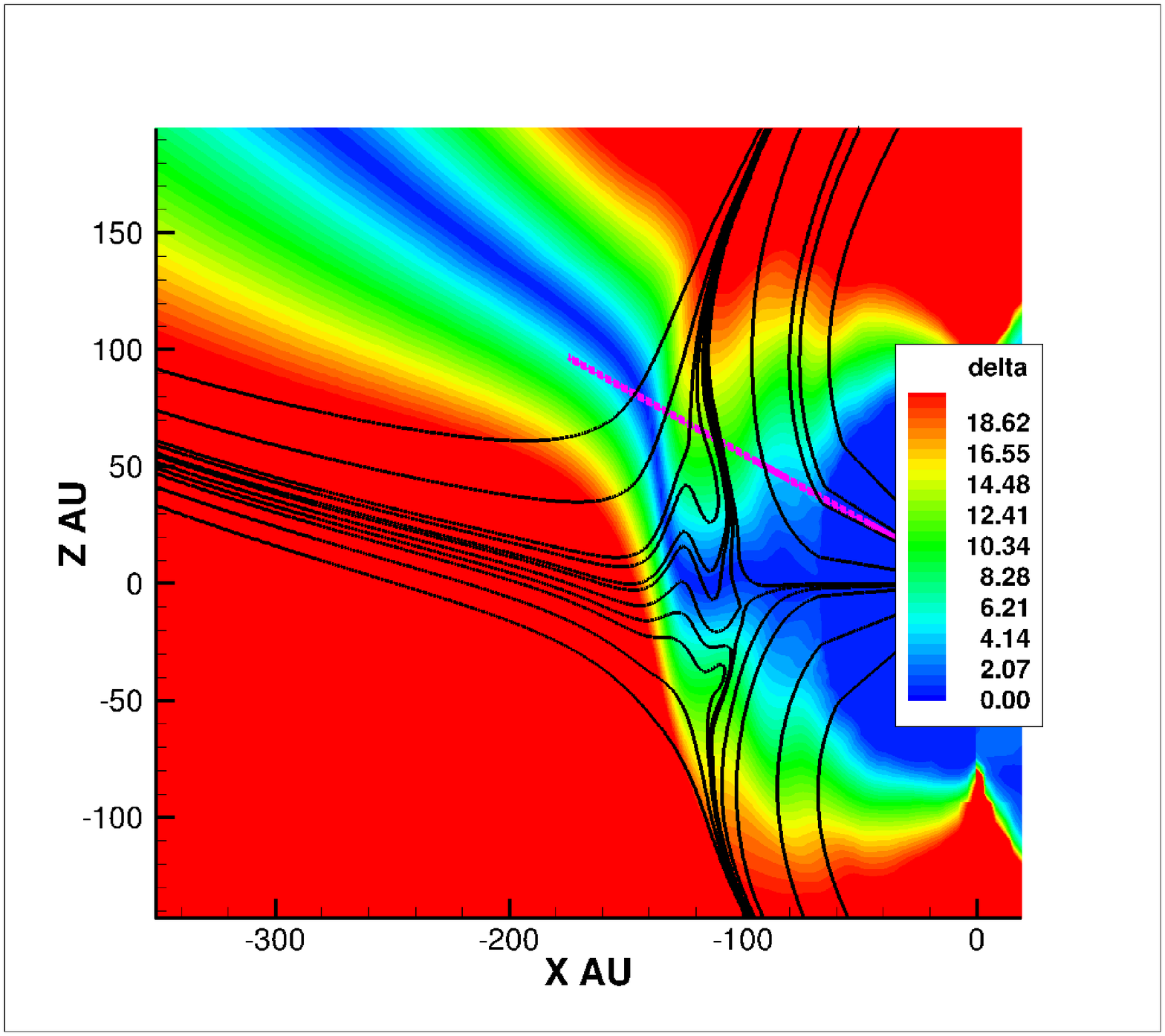}     
\caption{Same as Figure 1 but with a varying latitudinal solar wind as in Provornikova et al. (2013).}
\label{figure5}
\end{figure}

Both V1 and V2 are close enough to the stagnation point so that for both spacecraft there is a region outside of the HP where the angle $\delta$ remains small. This result is insensitive to the original orientation of $B_{ISM}$. Only far outside of the HP does the $B_{ISM}$ twist away from the Parker-like field direction. 

Far from the region of small $\delta$, how much of a twist of the magnetic field will V1 measure? Far from the HP, $B_{N}$ and the angle $\delta$ are zero at a certain angle from the solar equator. This angle is given by the orientation of $B_{ISM}$ as $\theta_{0}=90^{\circ}-tan^{-1} (1/(sin\beta tan\alpha)$, where $\alpha$ is the angle between the $B_{ISM}$ and the X- axis (approximately $\alpha_{BV}$) and $\beta$ is the angle between the solar equator and the BV plane). As shown by (Zieger et al. 2013) a slow bow shock can form ahead of the HP. As the interstellar magnetic field goes through a slow bow shock, the angle $\delta$ slightly changes. But, in any case this angle will be close to $\alpha_{BV}$. As argued in (Opher et al. 2009; Izmodenov et al. 2009) $\alpha_{BV}$ should be between $10^{\circ}-30^{\circ}$. This angle is very similar to the latitude of V1 ($30^{\circ}$ above the solar equator). The implication is that far from the HP the location where $B_{N}$ and $\delta$ are zero is nearly along the V1 trajectory. Therefore as V1 adventures farther away from the HP, $\delta$ will increase slightly but remain small. This increase will be stronger in cases with higher angle $\alpha_{BV}$ (as in the case where $\alpha_{BV} =45^{\circ}$; Figure 2b). The magnetic field direction at V2 will be very different. The magnetic field will exhibit a much larger twist, corresponding to a much higher values of $\delta$.  
 
 \section{Concluding Remarks}
 
These results suggest that the solar magnetic field plays a crucial role in controlling the draping of the interstellar magnetic field outside of the HP. Regardless of the orientation of $B_{ISM}$ the magnetic field twists to a Parker-like orientation just outside of the HP. The implication is that for neither V1 nor V2 can a strong magnetic field rotation out of the plane of the Parker spiral be used as a marker for the crossing of the HP.  On the other hand, we do expect some rotation in the field direction (or a change in the angle $\delta$) across the HP. Therefore, the several particle intensity dropouts detected by V1 from days 210-270, 2012 where there was no significant change in the direction of the magnetic field (Burlaga et al. 2013) cannot correspond to HP crossings. Our interpretation (Swisdak et al. 2013) is that the dropouts correspond to the separatrices of large-scale magnetic islands that form on the HP where the flux of heliospheric particles from the HS to the local interstellar medium (LISM) is suppressed. In this interpretation V1 crossed the HP on day 209 (when a current layer was crossed) and it has been measuring $B_{ISM}$ since that time. The angle $\delta$ reported during the subsequent period (Burlaga et al. 2013) is steady and around $14^{\circ}$, which is consistent with the results of our simulations in the region outside of the HP. Only after some distance from the HP will the spacecraft measure a substantial twist in the field, although in the case of V1 this twist is expected to be modest.

\acknowledgments
The authors would like to thank the staff at NASA Ames Research Center for the use of the Pleiades supercomputer. The authors acknowledge the support of the NASA-Voyager Guest Investigator grant NNX13AE04G to Boston University and NSF grant AGS-1202330 to the University of Maryland.

\end{document}